\pdfoutput=1
\documentclass[aps,pra,superscriptaddress,twocolumn,showkeys,10pt,nofootinbib]{revtex4-1}

\usepackage[pdftex]{graphicx}
\usepackage{amssymb, amsfonts, amsmath}
\usepackage{hyperref}
\usepackage[caption=false,singlelinecheck=false]{subfig}
\usepackage{multirow}

\begin{document}
\title[Scaling behavior of topologically constrained rings]%
{Scaling behavior of topologically constrained polymer rings in a melt}
\date{\today}

\author{Benjamin Trefz}
\affiliation{Johannes Gutenberg University Mainz, Department of Physics, Staudingerweg 7, 55128 Mainz, Germany}
\affiliation{Graduate School Material Science in Mainz, Staudinger Weg 9, 55128 Mainz, Germany}
\author{Peter Virnau}
\affiliation{Johannes Gutenberg University Mainz, Department of Physics, Staudingerweg 7, 55128 Mainz, Germany}

\begin{abstract}
Large scale molecular dynamics simulations on graphic processing units (GPUs) 
are employed to study the scaling behavior of ring polymers with various  
topological constraints in melts. Typical sizes of rings 
containing $3_1$, $5_1$ knots and catenanes made up of two unknotted rings scale 
like $N^{1/3}$ in the limit of large ring sizes $N$. This is consistent with the 
crumpled globule model and similar findings for unknotted rings.
For small ring lengths knots occupy a significant fraction of the ring. The 
scaling of typical ring sizes for small $N$ thus depends on the particular knot 
type and the exponent is generally larger than $0.4$.
\end{abstract}


\maketitle 

\section{Introduction}
Understanding the properties of ring polymer melts is one of the remaining major 
challenges in theoretical and experimental polymer physics 
\cite{rubinstein1986dynamics, cates1986conjectures, muller96, muller00, 
kapnistos2008unexpected, rosa2011structure, halverson11-1, halverson11-2, 
halverson2012rheology, lieberman2009comprehensive}. The topic became even more 
popular when ring melts were proposed as a model for DNA organization in a cell 
nucleus \cite{grosberg1993crumpled}: Chromatin fibers are highly packed and 
occupy ``territories'' \cite{cremer2001chromosome} just like unconcatenated 
rings in melts. \cite{rosa2011structure, halverson11-1, halverson11-2, 
dorier2009topological}.

Unlike linear polymers, rings cannot change their topological constraints once 
they are created, as bond crossings are not allowed. Thus, their behavior 
differs substantially from open chains. In a melt of open polymers the typical 
size of a chain scales like $R_{g}~\propto~N^{\frac{1}{2}}$ \cite{gennes} for large $N$, 
where $N$ stands for the number of monomers in the chain. In rings three 
different scaling regimes are present: For small chain lengths the radius of 
gyration scales with $R_{g}~\propto~N^{\frac{1}{2}}$, in an intermediate regime 
with $R_g \propto N^{\frac{2}{5}}$, while for long chain lengths a 
proportionality of $R_g \propto N^{\frac{1}{3}}$ \cite{cates1986conjectures, 
muller96, halverson11-1, halverson11-2, reith11gpu, vettorel09} will be reached. 
This asymptotic regime is generally associated with the crumpled globule concept 
from Grosberg et al. \cite{grosberg88} which states that the polymer ring 
collapses to a sphere-like object.

Although knots are seldom considered in theoretical derivations of scaling laws 
they become abundant in polymers and DNA for long chain lengths 
\cite{frisch61,delbruck62}. Even in relatively short proteins, knots have been 
reported \cite{mansfield, taylor00, lua06, virnau-07, virnau-11, potestio10, 
boelinger, nureki04, kolesob, mallam, jamroz2014knotprot} and also created 
artificially \cite{king10}. Topoisomerases are known to remove \cite{liu80} or 
create \cite{dean85} knots in DNA, which could otherwise inhibit transcription 
and replication. Viral DNA is known to be highly knotted inside capsids 
\cite{arsuaga02, arsuaga05, marenduzzo03, marenduzzo10, reith2}. Artificial 
knots have been tied in single DNA molecules with optical tweezers and 
dynamics have been studied both experimentally \cite{bao03} and with computer 
simulations \cite{vologodskii06}. Even though most of these examples are open 
chains and thus not knotted in a strict mathematical sense \cite{millett13}, where 
knots are only defined in closed curves, they nevertheless raise fundamental 
questions and challenge our understanding of topics as diverse as DNA ejection 
\cite{matthews09} and protein folding \cite{sulkowska09}.
 
In this paper we aim to combine these two topics and investigate the scaling 
behavior of topologically constrained polymer rings. To this end we compare a 
melt of unconcatenated, unknotted rings with their knotted or concatenated 
counterparts. 

In section 2 we describe the polymer model, simulation details, and the induced 
topological constraints. In section 3 we present our results for the scaling 
behavior 
and give a qualitative description of how knot sizes change for various chain 
lengths. Section 4 presents our conclusions.

\section{Model and Simulation Techniques}
We simulate $200$ ring polymers at a density of $\rho = 0{.}85 \sigma^{-3}$ in a 
simulation box with periodic boundary conditions. Chain lengths $N$ vary from 
$100$ monomers to up to $3{,}200$ monomers per polymer so the simulation box 
contains between $20{,}000$ and $640{,}000$ monomers. The temperature is set to 
$T=1{.}0\epsilon / k_{B}$ and is controlled with a Langevin thermostat. We use 
the open source software HooMD blue \cite{hoomd} and simulate each system on a 
Geforce GTX480 graphics card.

For each system size $2 \cdot 10^{9}$ MD steps with a time step of $\Delta t = 
0{.}01$ are simulated. The polymer model is based on \cite{halverson11-1}:
%\scriptsize
\begin{equation}
    U_{\textrm{lj}}(r_{ij}) = \left \{ 
				\begin{array}{l l} 
				4 \epsilon \left[ (\sigma / r_{ij})^{12} - (\sigma / r_{ij})^{6} \right]+\epsilon, &r_{ij} \le 2^{1/6} \sigma \\
				0, 										   &r_{ij}  >  2^{1/6} \sigma 
				\end{array} \right . \label{eq:lj}
\end{equation}
\begin{equation}
U_{\textrm{fene}}(r_{ij}) = \left \{ 
				\begin{array}{l l} 
				- 0.5 \ k \ R_{0}^{2} \ \ln \left[ 1 - (r_{ij} / R_{0})^{2} \right], 		& r_{ij} <   R_{0}	 	    \\
				\infty, 									& r_{ij} \ge R_{0} 
				\end{array} \right . \label{eq:fene}    
\end{equation}
\begin{equation}
U_{\textrm{angle}}(\theta_{i}) 	= \frac{1}{2} k_{\theta} \left( \theta_{i} - \pi \right)^{2} \quad , \label{eq:angle}    
\end{equation}
%\normalsize
with $k = 30\ \epsilon / \sigma^{2}$, $R_{0} =1.5\ \sigma$, $k_{\theta}=1.5\ 
\epsilon$. In contrast to \cite{halverson11-1} we employ the harmonic potential 
implemented 
in HooMD. Nevertheless, we expect that the results are very similar to 
\cite{halverson11-1} as equation~\ref{eq:angle} is essentially the second order 
Taylor expansion of the angular potential used by \cite{halverson11-1}. 
Therefore, the entanglement length of this model system is expected to be around 
$N_e \approx 28$ \cite{halverson11-1}.

\begin{figure}[!htb]
  \centering
  \includegraphics[keepaspectratio, width=8.3cm]{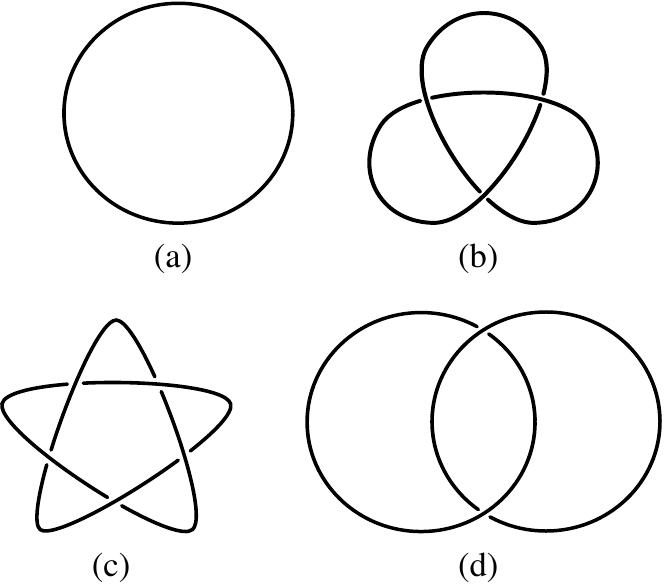}
  \caption{Schematic drawings of the examined topologies: (a) unknot (b) trefoil ($3_1$),  
(c) cinquefoil ($5_1$), and (d) catenane.}
  \label{pic:schematics}
\end{figure}
To investigate the influence of topology on the scaling of ring sizes in a melt 
we looked at various topologically constrained ring melts (see 
figure~\ref{pic:schematics} for schematic drawings). We study unknotted rings 
(figure~\ref{pic:schematics}(a)), knotted rings with a trefoil ($3_1$) knot 
(figure~\ref{pic:schematics}(b)) and with a cinquefoil ($5_1$) knot 
(figure~\ref{pic:schematics}(c)), and concatenated rings (catenane) 
(figure~\ref{pic:schematics}(d)). The model we use does not allow for bond 
crossings so that the initial constraints will be present at all times. To 
confirm this, we calculate the Alexander polynomial \cite{virnau10} for each 
knotted ring every $10^6$ MD steps. Additionally, we check that there are no 
bond crossings between the polymers via a primitive path analysis for all 
topologies \cite{everaers2004rheology}.
\begin{figure}[!htb]
  \centering
  \includegraphics[keepaspectratio, width=8.3cm]{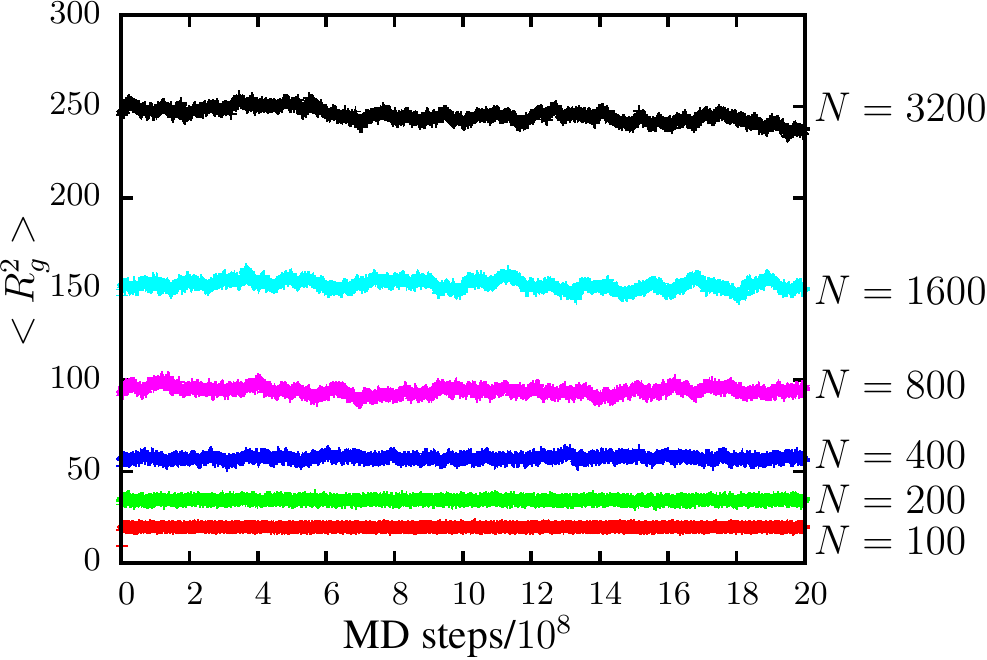}
  \caption{Production runs to determine the radius of gyration for unconstrained 
ring melts. The ring length N varies from $N=100$ (bottom) to $N=3{,}200$ (top). 
}
  \label{pic:equil}
\end{figure} 
To obtain the initial state we start with an already equilibrated configuration 
for a smaller ring length and put a monomer between each neighboring monomers on 
a polymer ring. We then rescale the system by a factor of $\sqrt[3]{2}$ in all 
dimensions and relax the system during a short ($3\cdot10^6$ MD steps) 
equilibration run. With this trick a prolonged equilibration period can be 
avoided. Nevertheless, each data point typically required several weeks of 
computation time on a single GPU. The largest systems ($N=3200$) ran for about 
$7$ months.

\section{Results}
To characterize the scaling behavior we calculate the squared radius of gyration 
${<R_{g}^{2}>}$ by averaging over all $200$ ring polymers every $10^6$ MD steps. 
Additionally, we monitor the squared mid-to-mid radius ${<R_{e}^{2}>}$, by 
averaging over the vector which connects the $i$th monomer to the $i+N/2$th 
monomer.

\begin{figure}[!htb]
  \centering
  \subfloat[]
    {\includegraphics[keepaspectratio, width=8.3cm]{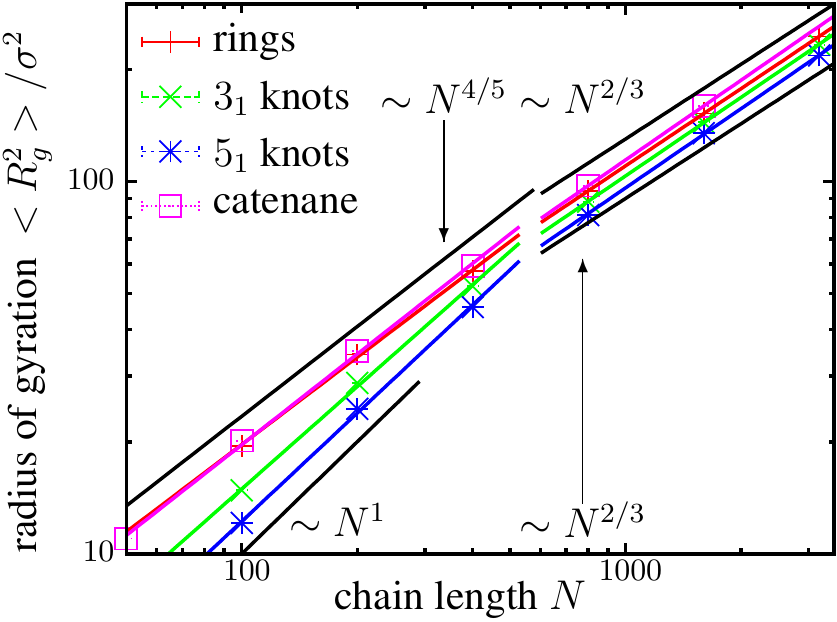} \label{pic:gyr} }\\
  \subfloat[]
    {\includegraphics[keepaspectratio, width=8.3cm]{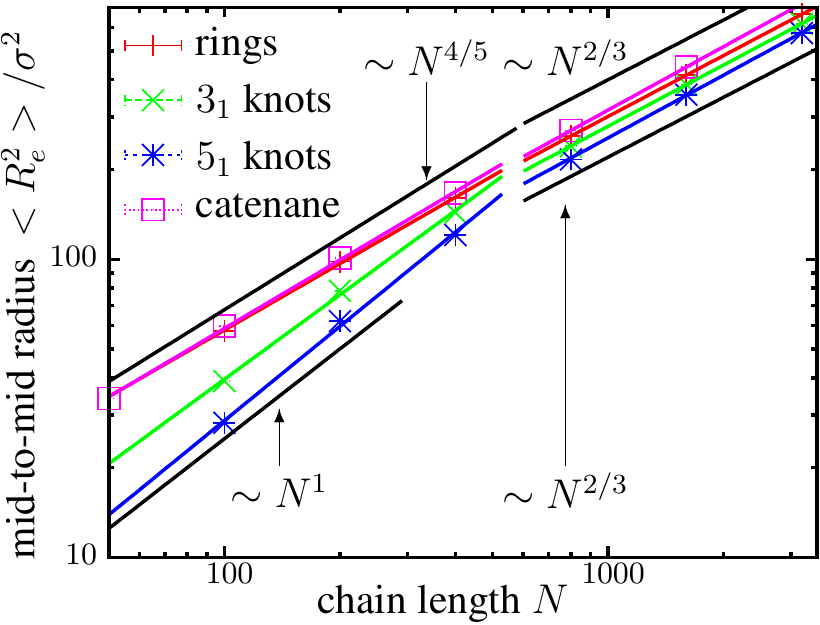} \label{pic:end}}
  \caption{(a) Scaling behavior of the squared radius of gyration for polymer 
rings with various topological constraints. (b) Scaling behavior of the squared 
mid-to-mid radius for polymer rings with various topological constraints.}
  \label{pic:radii}
\end{figure}

By plotting the average squared radius of gyration or the mid-to-mid radius 
against the chain length $N$ (see figure~\ref{pic:radii}) we can obtain the 
scaling behavior by fitting a power law \footnote{It should be noted that our 
data only spans a bit more than a decade and exponents are typically obtained by 
fitting only three points.} of the form $R^{2}(N) = a \cdot N^{2 \nu}$. 

\begin{table}[!htb]
\begin{tabular}{|l|l|l|l|l|}
\hline 
\multirow{2}{*}{Topology} 	& \multicolumn{2}{|c|}{$R_{g}$} & \multicolumn{2}{|c|}{$R_{e}$} \\
\cline{2-5}
 ~		& $2 \nu_1$	& $2 \nu_2$ 	& $2 \nu_1$ 	&  $2 \nu_2$	\\
\hline
rings		& $0.779$	& $0.689$	& $0.740$ 	&  $0.679$ 	\\
$3_1$ knots	& $0.912$ 	& $0.700$	& $0.940$ 	&  $0.683$ 	\\
$5_1$ knots	& $0.966$	& $0.706$	& $1.050$ 	&  $0.699$	\\
catenanes 	& $0.809$	& $0.711$	& $0.762$ 	&  $0.707$	\\
\hline
\end{tabular}
\caption{Exponents for the chain length dependency of the radius of gyration and
the mid-to-mid radius obtained by fitting the function $R^{2}(N) = a \cdot N^{2 \nu_1}$ 
to the data points with $N \le 400$ and $R^{2}(N) = b \cdot N^{2 \nu_2}$ for the 
region $N \ge 800$ as shown in figure~\ref{pic:gyr}. For unconstrained rings the 
expected exponents are $2 \nu_1 = 0.8$ and $2 \nu_2 = \frac{2}{3}$. }
\label{tab:fitpar}
\end{table}

\begin{figure}[!htb]
  \centering
  \includegraphics[keepaspectratio,width=8.3cm]{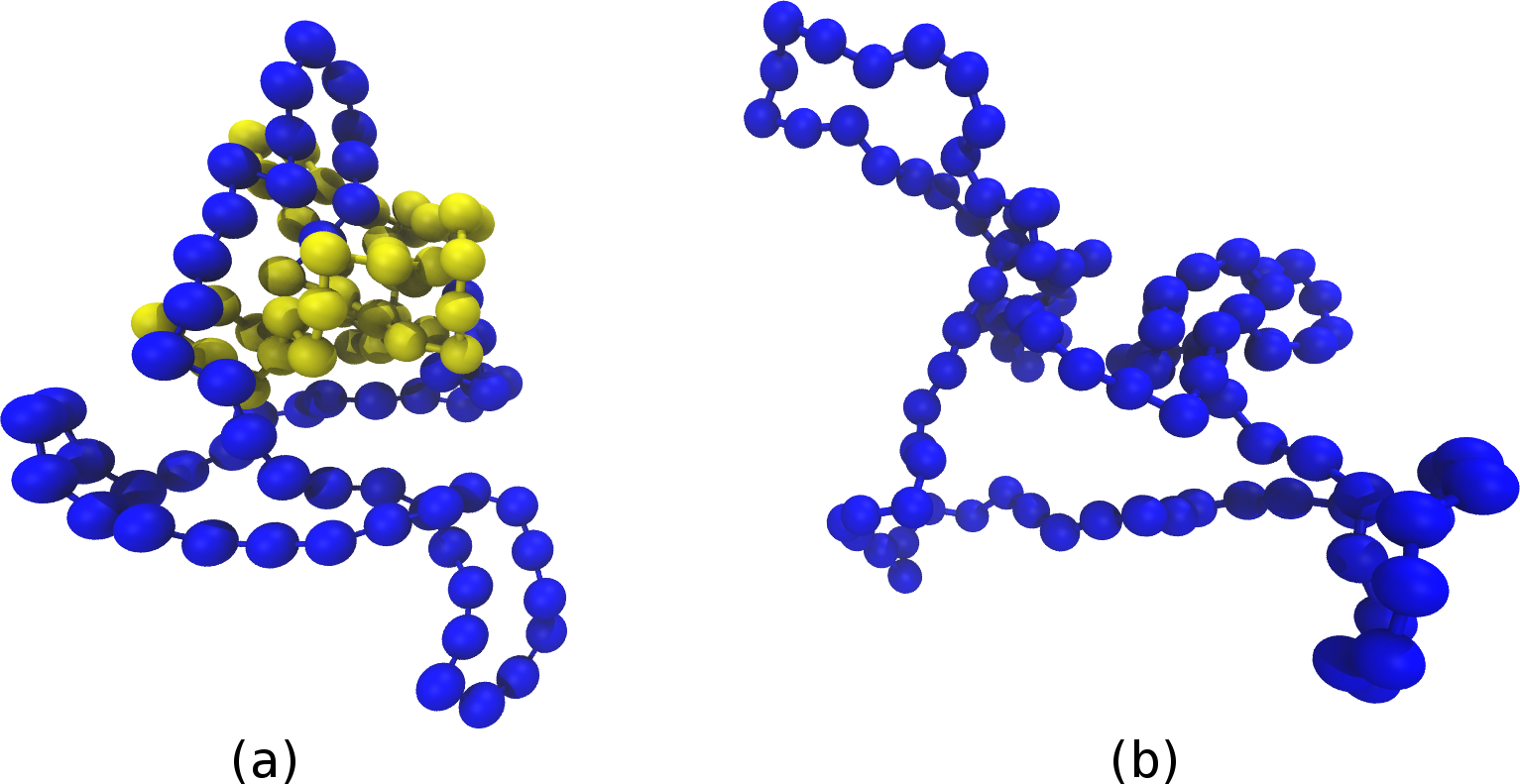}
  \caption{Typical snapshots: (a) $5_1$ knot with ring size $N=100$, the knotted 
part is colored in red. (b) Ring with the same ring size, but without constraint.
Configurations were chosen to have typical radii of gyration. For small $N$
the knotted part leads to smaller, more compact conformations.}
  \label{pic:ringknot}
\end{figure}

\begin{figure}[!htb]
  \centering
  \includegraphics[keepaspectratio,width=8.3cm]{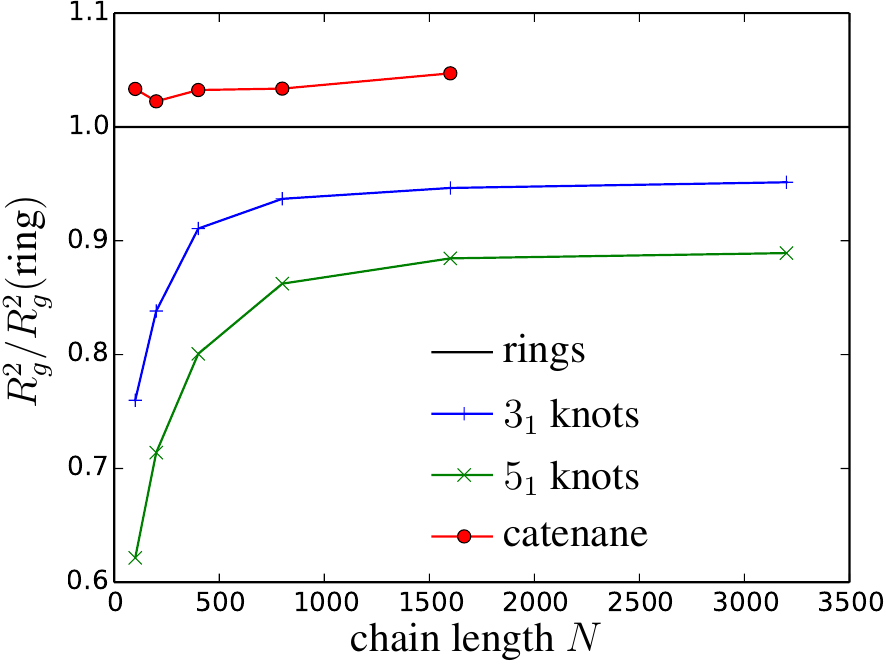}
  \caption{Ratio of radius of gyration for systems with topological constraint 
and the corresponding radius of the unconstrained ring melt. The influence of a 
knot on a ring is lessened with increasing chain length $N$. The concatenated 
rings (catenane) are slightly larger than the free rings, as they have to stay 
in pairs of two.}
  \label{pic:knotgrowth}
\end{figure}

For all topologies we observe a scaling exponent compatible with $\nu_2 \approx 
\frac{1}{3}$ as predicted by the crumpled globule model \cite{grosberg88} (see 
table~\ref{tab:fitpar}). For small systems, however, knots occupy a significant 
fraction of the polymer (see figure~\ref{pic:ringknot}) resulting in a denser 
configuration with a smaller radius of gyration. This effect levels off (see 
figure~\ref{pic:knotgrowth}) around the transition to the crumpled globule 
states (between 400 and 800 monomers). Hence, the effective exponent $\nu_1$ 
exceeds the predicted value of $0.4$ for unconstrained rings. The catenanes 
behave very similar to the unconstrained rings, and the radius of gyration 
scales with the same exponent (see figure~\ref{pic:radii}). Still, they are 
slightly larger than their unconstrained counterparts (see 
figure~\ref{pic:knotgrowth}) for all ring sizes $N$. This is expected, as the 
catenanes have to stay in pairs of two and can thus not relax as easily as the 
unconstrained rings. 

For knotted rings figure~\ref{pic:knotgrowth} suggests that the influence of the 
knot on the total size decreases as a function of $N$. Unfortunately, it is very 
difficult to determine sizes of knots in rings 
\cite{tubiana2011probing,rosa2011structure} and our preliminary findings are not 
completely conclusive. To determine knot sizes we have followed a so called 
top-down approach \cite{rosa2011structure}: We cut the chain at a random point 
and remove beads successively until the knot disappears. The remaining beads are 
essential and determine the knot size. However, if we cut inside the knot the 
analysis does not yield meaningful results. The closure may also lead to 
additional distortions. By considering multiple random starting points and 
analyzing only the subset of data which yields consistent sizes for at least 
half of the starting points, we have determined the most likely knot size as 
$40$ monomers for the trefoil knot and $51$ monomers for the $5_1$ knot for ring 
size $N=100$. For larger rings the most likely size is similar, but the 
distribution has a larger tail.

\section{Conclusion}
In conclusion, we have performed large scale molecular dynamics simulation on 
graphic processing units (GPUs) to study the scaling behavior of unknotted ring 
polymers, 
knotted rings with a trefoil ($3_1$) knot and with a cinquefoil ($5_1$) knot, 
and concatenated rings in a melt. For large $N$ the rings scale with roughly 
$N^{1/3}$ for all topological constraints, and the transition takes place at 
about the same ring sizes. These findings for large ring sizes are consistent 
with the crumpled globule model. For small ring lengths knots occupy a 
significant fraction of the ring. The scaling of typical ring sizes for small 
$N$ thus depends on the particular knot type and the exponent is generally 
larger than $0.4$.

\section{Acknowledgment}
B.T. and P.V. would like to acknowledge the MAINZ Graduate School of Excellence 
for financial support and the ZDV Mainz for computational resources.

%\section*{Literature}
\bibliographystyle{unsrt}
\bibliography{./literature}

\end{document}